\documentclass{aastex631}

\usepackage{float}

\begin{document}

\title{Surprise non-detection of Upsilon Andromedae b with MIRC-X and MYSTIC at the CHARA Array}

\author[0000-0002-3003-3183]{Tyler Gardner}
\affiliation{Astrophysics Group, Department of Physics \& Astronomy, University of Exeter, Stocker Road, Exeter, EX4 4QL, UK}
\author[0000-0002-3380-3307]{John D. Monnier}
\affiliation{Astronomy Department, University of Michigan, Ann Arbor, MI 48109, USA}
\author[0000-0001-6017-8773]{Stefan Kraus}
\affiliation{Astrophysics Group, Department of Physics \& Astronomy, University of Exeter, Stocker Road, Exeter, EX4 4QL, UK}
\author[0000-0003-3963-9672]{Emily Rauscher}
\affiliation{Astronomy Department, University of Michigan, Ann Arbor, MI 48109, USA}
\author[0000-0003-0217-3880]{Isaac Malsky}
\affil{Jet Propulsion Laboratory, California Institute of Technology, Pasadena, CA 91109, USA}
\author[0000-0002-0493-4674]{Jean-Baptiste Le Bouquin}
\affiliation{Institut de Planetologie et d'Astrophysique de Grenoble, Grenoble 38058, France}
\author[0000-0002-2208-6541]{Narsireddy Anugu}
\affiliation{The CHARA Array of Georgia State University, Mount Wilson Observatory, Mount Wilson, CA 91203, USA}
\author{Sorabh Chhabra}
\affiliation{Astrophysics Group, Department of Physics \& Astronomy, University of Exeter, Stocker Road, Exeter, EX4 4QL, UK}
\author[0009-0005-8088-0718]{Isabelle Codron}
\affiliation{Astrophysics Group, Department of Physics \& Astronomy, University of Exeter, Stocker Road, Exeter, EX4 4QL, UK}
\author[0000-0001-9764-2357]{Claire L. Davies}
\affiliation{Astrophysics Group, Department of Physics \& Astronomy, University of Exeter, Stocker Road, Exeter, EX4 4QL, UK}
\author[0000-0002-1788-9366]{Noura Ibrahim}
\affiliation{Astronomy Department, University of Michigan, Ann Arbor, MI 48109, USA}
\author[0000-0001-9745-5834]{Cyprien Lanthermann}
\affiliation{The CHARA Array of Georgia State University, Mount Wilson Observatory, Mount Wilson, CA 91203, USA}
\author[0000-0001-5415-9189]{Gail Schaefer}
\affiliation{The CHARA Array of Georgia State University, Mount Wilson Observatory, Mount Wilson, CA 91203, USA}
\author[0000-0001-5980-0246]{Benjamin R. Setterholm}
\affiliation{Max-Planck-Institut für Astronomie, Heidelberg, Germany}



\begin{abstract}

Ground-based long baseline interferometry is a powerful tool for characterizing exoplanets which are too close to their host star to be imaged with single-dish telescopes. The CHARA Array can resolve companions down to 0.5 milli-arcseconds, allowing us in principle to directly measure the near-infrared spectra of non-transiting ``Hot Jupiter" exoplanets. We present data taken with the MIRC-X and MYSTIC instruments at the CHARA Array on the Hot Jupiter Upsilon Andromedae b. By resolving the star-planet system, we attempt to directly detect the flux from the planet. We describe our self-calibration methods for modeling systematics in the closure phase data, which allows us to reach sub-degree precision. Through combining multiple nights of data across two MIRC-X runs in 2019 and 2021, we achieved a very tentative detection of Ups And b in the H-band at a planet/star contrast of 2-3$\times$10$^{-4}$. Unfortunately, we cannot confirm this detection with 2021 MYSTIC data in the K-band, or in a 2023 joint MIRC-X and MYSTIC dataset. We run updated global circulation models and create post-processed spectra for this planet, and report the resulting model spectra in H- and K-bands as a function of orbital phase. We then run planetary injection tests to explore H/K-band contrast limits, and find that we can confidently recover planets down to a planet/star contrast of 1--2$\times$10$^{-4}$. We show that we are probing contrasts fainter than predicted by the model, making our non-detection surprising. We discuss prospects for the future in using this method to characterize companions with interferometry.

\end{abstract}

\keywords{exoplanets: direct detection, interferometry}


\section{Introduction}\label{intro}
Hot Jupiters are ideal targets for constraining exoplanet atmosphere models due to their favorable planet-to-star flux ratios at infrared wavelengths compared to smaller or wider orbit exoplanets. To date, detections of transmitted starlight (starlight passing through planet atmospheres) or emitted light from planets have been made for dozens of transiting Hot Jupiter exoplanets. Studies of transiting planets have yielded an abundance of information on their atmospheric characteristics through transmission spectroscopy \citep[e.g.][]{nikolov2014,kreidberg2014,samra2023}. Orbital phase variations of strongly irradiated Hot Jupiters have been detected from space \citep[e.g.][]{cowan2012,stevenson2014,parmentier2018,may2018}, but previous studies are mostly restricted to transiting planets which limits the number of targets for which we can measure variations in spectra around a planet's full orbit. High spectral resolution (R$\sim$100,000) and high SNR radial velocity observations have also detected signatures from CO from a few non-transiting Hot Jupiters including $\tau$ Boo b and 51 Peg b \citep{rodler2012,brogi2013}. These methods were used to detect water vapor in the atmosphere of the target of interest of this paper, Upsilon Andromedae b \citep{piskorz2017}. This work measured a planet orbital inclination of 24$\pm$4$^\circ$ and planet mass of 1.7$^{+0.33}_{-0.24}$ M$_{Jup}$. However, the results are now in doubt after follow-up injection tests suggested that the planets should not have been detectable with the available data \citep{buzard2021}. 

Despite all the exciting progress on observing Hot Jupiters, key questions, like the diversity of circulation efficiencies and thermal inversion in some planets, are still not fully understood \citep{fortney2021}. We develop an emerging method for characterizing Hot Jupiter planets with long-baseline interferometry. Interferometric closure phases constrain the asymmetry in the brightness distribution of the source, making it a powerful observable for companion detection. High angular resolution interferometers can in principle be used to resolve the planet from the star and measure a low-resolution spectrum of exoplanets over the full orbital cycle. This full orbital monitoring would be an advantage over the snapshots provided by transit spectroscopy at primary and secondary eclipse, or over phase curves that blend the star and planet light together. In addition, an interferometric detection of a Hot Jupiter system would naturally reveal the orbit and mass of the exoplanet for non-transiting planets (similar to how interferometers target high contrast binary systems for mass measurements). This method would also provide a way to study 3D temperature variations across the planet, as well as the polar regions of non-transiting Hot Jupiters. Detection and characterization with long-baseline interferometry would provide crucial information for understanding planet properties of Hot Jupiter systems. 

Ground-based long baseline interferometry is already becoming a valuable tool in the field of exoplanet science. The ExoGRAVITY team, for example, is using the GRAVITY instrument at VLTI to directly detect and measure spectra and orbits of planets separated by $\sim$50--200 milli-arcseconds (mas) from their host star \citep[e.g]{nowak2020,nasedkin2024,winterhalder2024,balmer2024}. This opens up a new parameter space for planet characterization, as these exoplanets are often within the diffraction limit of single telescopes. The methods of the ExoGRAVITY team involve placing a fiber on the known planet position and tracking fringes on the host star, though this ``dual field" mode currently restricts characterization of planets closer than $\sim$50 mas. The CHARA Array in the northern hemisphere is composed of six telescopes separated by baselines up to 330 meters. In the H-band, this results in an angular resolution of about 0.5 mas. This is enough resolution to resolve a number of Hot Jupiter systems from their host star. If the high contrast can be achieved, this opens up a whole new regime of exoplanets for characterization. Single-field interferometry, which utilizes the full resolution power of the method, is challenging for exoplanet work because of the blended light from the planet and star, leading to noise from the many stellar photons \citep{lacour2023}. In addition, a planet with flux 10$^{-4}$ fainter than its star produces a phase signal on the order of 0.01$^{\circ}$, which is extremely challenging for current instruments. We started a project attempting to achieve this phase precision with the MIRC-X and MYSTIC instruments at the CHARA Array to measure low-resolution spectra of non-transiting Hot Jupiter planets, nicknamed Project PRIME (PRecision Interferometry with MIRC-X for Exoplanets). 

The pioneering work for the PRIME project was carried out by \citet{zhao2011}. In these efforts, the MIRC instrument at the CHARA Array was used to target the Hot Jupiter Upsilon Andromedae b. Strict upper limits were placed on the planetary flux at a level of a 5-8 x 10$^{-4}$, only a factor of $\sim$3 away from detecting the expected flux level of the planet in H-band. Since this work, the MIRC instrument has been updated from a 4-telescope combiner to 6-telescopes. It has also undergone multiple optics and detector upgrades to become the current MIRC-X instrument \citep{anugu2020}. Recently, the MYSTIC instrument has also been comissioned at the CHARA Array to simultaneously record K-band data with MIRC-X \citep{setterholm2023}. These upgrades have motivated a new attempt to detect the flux from non-transiting Hot Jupiter exoplanet Ups And b. If such a detection can be achieved, this would demonstrate a new method for characterizing close-in planets. Along with being a valuable tool for studying non-transiting Hot Jupiters, such a method would also become useful in the era of Gaia. Gaia discovered hundreds of thousands of new companions in DR3 \citep{gaia_dr3}, many of which are too close-in to their host star to be characterized with non-interferometric techniques. Future data releases will include more substellar and planetary mass companions, and long baseline interferometry can be used for these high-contrast companions to measure low-resolution spectra in the near-infrared wavelengths, if we are able to demonstrate the high contrast required. 

This paper is organized as follows. Section \ref{observations} describes our observational setup for project PRIME with the MIRC-X and MYSTIC instruments at the CHARA Array to detect Ups And b. In Section \ref{analysis} we describe our data reduction routines, which includes a self-calibration method to achieve the sub-degree closure phase precision needed for detecting Hot Jupiters. We show our results in Section \ref{results}, which includes a very tentative detection with the MIRC-X instrument in a 2019Sep run, although we are unable to confidently confirm this detection in follow-up runs. In Section \ref{spectra} we present updated model spectra for Ups And b using our best guess at the orbital inclination and planet mass. We also run planetary injection tests to derive contrast limits for our instruments, showing that we are probing deeper contrasts than those predicted by the models. We give concluding remarks and prospects for the future of this method in Section \ref{conclusions}.

\section{Observations}\label{observations}

H-band data for PRIME are taken with the Michigan InfraRed Combiner-Exeter (MIRC-X) instrument at the CHARA Array. The CHARA Array has six telescopes with baselines up to 330 meters, giving it the longest baselines in the world of any optical interferometer \citep{tenbrummelaar2005}. The MIRC-X instrument combines all six telescopes, with angular resolution in the H-band down to 0.5 mas. The original MIRC instrument is described in detail in \citet{Monnier2006}, and the upgraded MIRC-X instrument is described in \citet{anugu2020}. We had three runs of MIRC-X data on the Hot Jupiter Ups And b in 2019 September (5 nights), 2021 October (4 nights), and 2023 October (5 nights). For each run, we attempted to cover a full orbit of the planet. Future publications will also include a fourth run (2022 October) with a Wollaston prism in the beam to model polarization behavior of the array (beyond the scope of this paper). All MIRC-X datasets were taken in H-grism mode with R$\sim$190. 

Normally, interferometric observations employ frequent pointings to calibrator stars (sources with known sizes). This allows one to calibrate losses in the visibility due to atmospheric or systematic effects. For our program we are attempting to detect the Hot Jupiter in the closure phase signal, which is immune to atmospheric effects. \citet{zhao2011} showed that the level of precision needed to detect Ups And b in the closure phase is at the $\sim$0.1 degree level in H-band. At this level, we detect systematic features in closure phase across the night which need to be corrected for. While pure phase disturbances should not directly affect the measured closure phases, we have identified potential sources of systematics that can corrupt closure phases in MIRC-X: 1) Flat field error and bad pixels can cause cross-talk between baselines when carrying out the Fourier Transform in an all-in-one combiner. We track to zero optical path differences (OPDs), so there can be a systematic that results that is seeing dependent. 2) The PSF of the spectrograph may not be the same for all baselines since the light from each baseline takes a different optical path through the system. This has the effect of having a slightly different effective wavelength per baseline, which couples into a closure phase error as the differential dispersion and OPD tracking changes in time during the night. 3) Birefringence introduces differential phase shifts between the horizonatal and vertical polarized fringes. The beam train becomes about 10-20\% polarized due to the large number of 45 degree reflections, thus introducing a small systematic as the birefringence changes. Though we did observe some calibrators on these nights, we find that the target data itself is best for modeling closure phase systematics (since Ups And is bright and we spent most of the night on-target, our closure phase precision is significantly better compared to shorter calibrator pointings). We describe our self-calibration routine in the next section. 

In our 2021 and 2023 October runs, we also have data taken with the MYSTIC instrument. MYSTIC takes data in the K-band, and can be used simultaneously with the MIRC-X instrument \citep{setterholm2023}. For K-band MYSTIC, we used the prism R$\sim$50 mode in 2021 and the prism R$\sim$100 mode in 2023. The rest of the observing setup was the same, since the data for both instruments were taken at the same time. All of the data used for this paper are public on the JMMC's database \footnote{\url{https://oidb.jmmc.fr/index.html}}. Since Ups And is a bright star (H=3.0) that is only resolved on the longest baselines (UD=1.097 mas; \citealt{zhao2011}), this is a relatively simple target to observe with MIRC-X/MYSTIC. The challenge of this program comes in achieving the SNR and closure phase precision to be able to detect the flux from the planet which is at an expected level of contrast approaching 10,000:1. In the following section, we describe our calibration method to achieve this precision in closure phase. 


\section{Data Analysis}\label{analysis}
\subsection{Reduction Pipeline}
The MIRC-X and MYSTIC instruments measure visibilities, differential phase, and closure phase of our targets. We use the MIRC-X pipeline (version 1.3.3) to produce OIFITS files for each night. This pipeline is described in \citet{anugu2020}, and maintained on Gitlab\footnote{\url{https://gitlab.chara.gsu.edu/lebouquj/mircx_pipeline}}. At this stage, we do not yet carry out any calibration of our target data, meaning that the closure phase and visibility that the pipeline measures will be raw (i.e. uncalibrated). Since we require extra cleaning and calibration steps for project PRIME, we want short integrations of closure phase at this step which will be averaged together at a later step. Hence we reduce our data with an oifits max integration time of 30 seconds, and a number of coherent integration frames (ncoh) of 10. We also apply the bispectrum bias correction of the pipeline, inspired by the methods of \citet{basden2004} and \citet{garcia2016}, with our procedure described in \citet{anugu2020}. 

After obtaining our 30-second chunks of closure phase measurements across a night, we then clean up our data with an IDL routine specially made for this program. This routine follows the principles described in \citet{monnier2012} for the MIRC pipeline, which was well-tested to flag and remove bad data and average together closure phase and visibility. This script then cleans the data by removing outliers and averages the closure phase measurements into 15-minute chunks. For MYSTIC low resolution mode, we now have clean datasets with 7--9 spectral channels for the R$\sim$50 mode. For our MIRC-X and MYSTIC data taken in higher resolution mode, we perform a median filter across wavelength to average our data from $\sim$34 spectral channels to 7--9 channels. At this step we now have cleaned and averaged closure phases for each night, though these measurements are still uncalibrated. In the next section we describe our self-calibration routine to remove trends in phase due to systematics.

\subsection{Self-Calibration for Achieving Precision Closure Phase}

\begin{figure}[H]
\centering
\includegraphics[width=\textwidth]{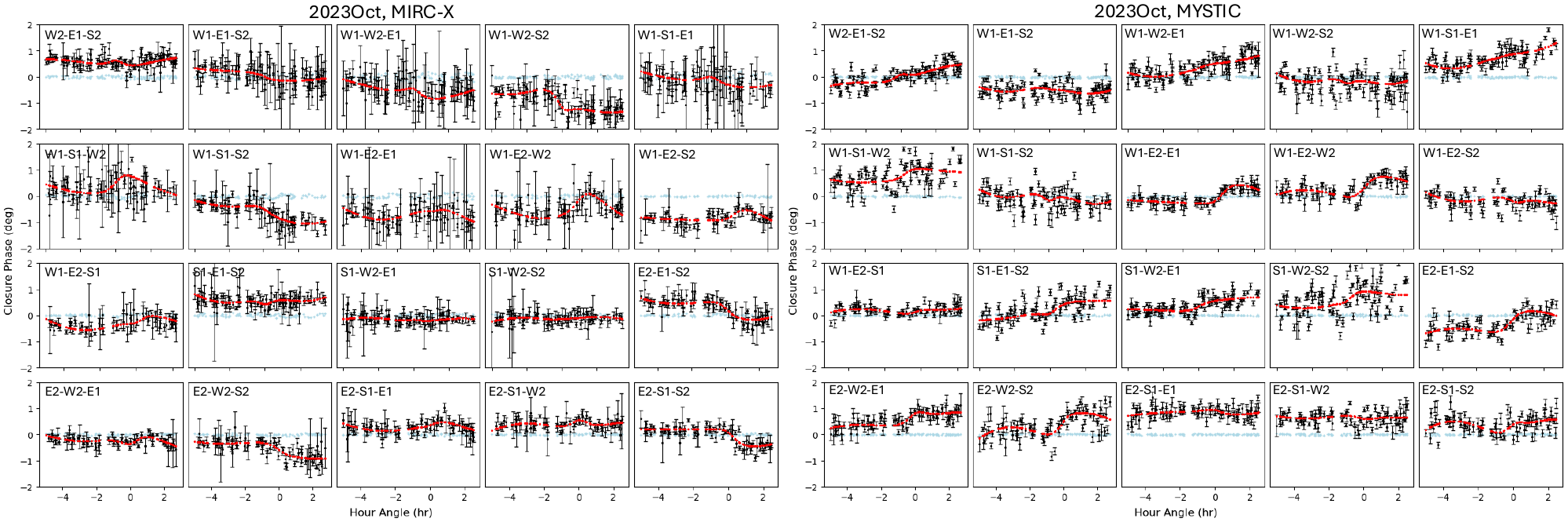}
\caption{ We plot the MIRC-X (left) and MYSTIC (right) closure phase for one spectral channel across hour angle during the 2023Oct run. Each box represents one of the 20 closing triangles of the CHARA Array, with the designated telescope names labeled. Though we are already looking at sub-degree features, the drifts and dips need to be calibrated out in order to reach the precision needed to detect the flux from a planet. The black data points represent the closure phase data, while the red points show our model of systematic trends. The light blue crosses show the expected planet signal at a planet-to-star contrast ratio of 2e-4.}
\label{trends}
\end{figure}

At this stage before additional calibration, our closure phase precision is already at the sub-degree level in standard deviation across time and triangles for both MIRCX and MYSTIC. However, we see systematic features in the closure phase which are likely not from the Hot Jupiter planet since the amplitude of these signals is significantly greater than the expected $\sim$0.1 degree signal from the planet. We also do not find any convincing companion fit to these closure phase feautures, suggesting they are instrumental trends. Figure \ref{trends} shows closure phase data across hour angle for one spectral channel during the run of 2023Oct (data is plotted for 2019Sep and 2021Oct in Figures \ref{trends_2019Sep} and \ref{trends_2021Oct} in the appendix). As can be seen, there are clear drifts with hour angle and some odd ``dips" in the closure phase signal. These need to be calibrated out in order to detect any signal from a Hot Jupiter planet. It is especially worthy to note that multiple triangles share a strong ``dip" feature just before the source transits (i.e. passes through its highest point in the sky). This potentially hints at a polarization issue which is affecting our data at a sub-degree level. Although it is outside the scope of this paper, our team recently submitted a paper that delves into the polarization behavior of these instruments in detail (Shuai et al., submitted).

\citet{zhao2011} described a calibration model which improved the precision of MIRC data on Ups And b. In that work, the authors used calibrator sources to model closure phase behavior as a function of altitude and azimuth of the target in the sky. However, the expected closure phase signal from the Hot Jupiter exoplanet does not follow these trends with altitude and azimuth. Hence, we can use the target data itself to model these systematic effects. This way we do not need to extrapolate the model into regions where there is no calibrator data, and we have higher precision data to model such effects since more time is spent on the bright, high SNR target star. Additionally, it means that we are able to integrate on-target all night which increases our coverage and signal for the extremely faint Hot Jupiter companion. Similar to \citet{zhao2011}, we use the following model to calibrate the systematic features in closure phase with altitude and azimuth:

\begin{equation}\label{cp_correction}
    \phi = a_0 + a_1*Az + a_2*Az^2 + a_3*Az*Alt + a_4*Alt + a_5*Alt^2 ,
\end{equation}

where $\phi$ is the closure phase signal due to systematic drifts, $a_n$ are fitted coefficients, and Az and Alt refer to the target azimuth and altitude. 

We use the model of Equation \ref{cp_correction} to fit for each spectral channel and closing triangle separately, and combine all the nights of a single run together to fit for the coefficients. Figure \ref{trends} also includes our model fit for the 2023Oct run. We are able to capture many of the drifts with our model of altitude and azimuth, which increases the closure phase precision for both instruments. Once we subtract out these self-calibrated systematics from the target closure phase data, we are ready to search for the flux from the Hot Jupiter planet. 

\subsection{Grid Search Routine to Detect Ups And b}
Once we have our final reduced, cleaned, and self-calibrated closure phase data, we can begin to search for the signal from the high contrast planet. To do so, we fit a Keplerian orbit model directly to the closure phases. All nights from a single run are fit simultaneously. The orbital elements for a planet/binary orbit consist of an orbital period $P$, a semi-major axis $a$, an inclination $i$, an eccentricity $e$, the longitude of periastron $\omega$, the position angle of the ascending node $\Omega$, and the time of passage through periastron $T$. For a given observation time these orbital elements will predict the position ($\Delta$x, $\Delta$y) of a companion relative to the primary star at the origin. The closure phase signal is then set by this position, along with a flux ratio $f = f_{star} / f_{planet}$, and a uniform disk (UD) size for the primary star and planet. We can therefoxre fit our data with the binary orbital elements, planet/star contrast, and uniform disk size of Ups And as free parameters. 

Ups And b has been characterized in RV, which provides many of the known orbital elements of the planet. \citet{mcarthur2010} published the most precise RV orbit of this planet, and hence we are able to use these values to inform our searches. We fix $P=4.617111$ days, $e=0.012$, $\omega=224.11^{\circ}$ (180$^{\circ}$ off from the RV value, as this parameter is defined differently for visual and RV orbits), and $T = 50033.55$ MJD from the fitted values to the RV data for Ups And b. These values were obtained 15 years ago and we can propagate the expected errors to determine if orbital uncertainties need to be addressed in our modeling.  We find that the expected phase error in 2024 is $\pm$ 0.003 (of a period).  This timing error corresponds to a maximum astrometric error of only 0.09mas, which is 1/10 of a fringe for our longest baselines, and thus can be ignored. In future work, we will incorporate more recent RV data to allow for a more precise ephemeris, but is not necessary for the work presented here. We also set the host star $UD = 1.097$ mas \citep{zhao2011}, and we keep the planet size unresolved at 0.1 mas. Our free parameters are then $a$, $i$, $\Omega$, and $f$. However, since we know the mass of the star (1.31$\pm$0.02 M$_{\odot}$, \citealt{piskorz2017}), the orbital period, and the distance to the system (13.48$\pm$0.04 pc, \citealt{gaia_dr3}) we also have a good value for the orbital semi-major axis: $a=4.41\pm0.05$ mas \citep{rosenthal2021}. We also note that \citet{piskorz2017} reported the orbital inclination of Ups And b to be 24$\pm$4$^{\circ}$. However, this value is not as reliable as the others, since it depended on the detection of water vapor in the atmosphere of the planet which \citet{buzard2021} calls into question based on simulations of these features which suggest that it should not have been detectable. To search for the planet we perform a brute force grid search over the parameters of $a$, $i$, and $\Omega$, letting the flux ratio $f$ vary as a free parameter at each step. We vary $a$ from 3--6 mas at a step size of 0.1 mas, $i$ from 0--180$^{\circ}$ at a step size of 3$^{\circ}$, and $\Omega$ from 0--360$^{\circ}$ at a step size of 2$^{\circ}$. These grid sizes were all confirmed to be able to recover a planet with injection tests described in Section \ref{mystic_injection}. We then use a Bayesian Information Criteria (BIC) to find potential solutions for the planet, described in the next section. We use the Python package \textit{lmfit} to perform our $\chi^2$ minimization at each step of the grid (\citealt{newville2016}, note that only the flux ratio varies as a free parameter at each step).

\clearpage
\section{Results}\label{results}

\subsection{Grid Search for Ups And b}
\begin{figure}[H]
\centering
\includegraphics[width=\textwidth]{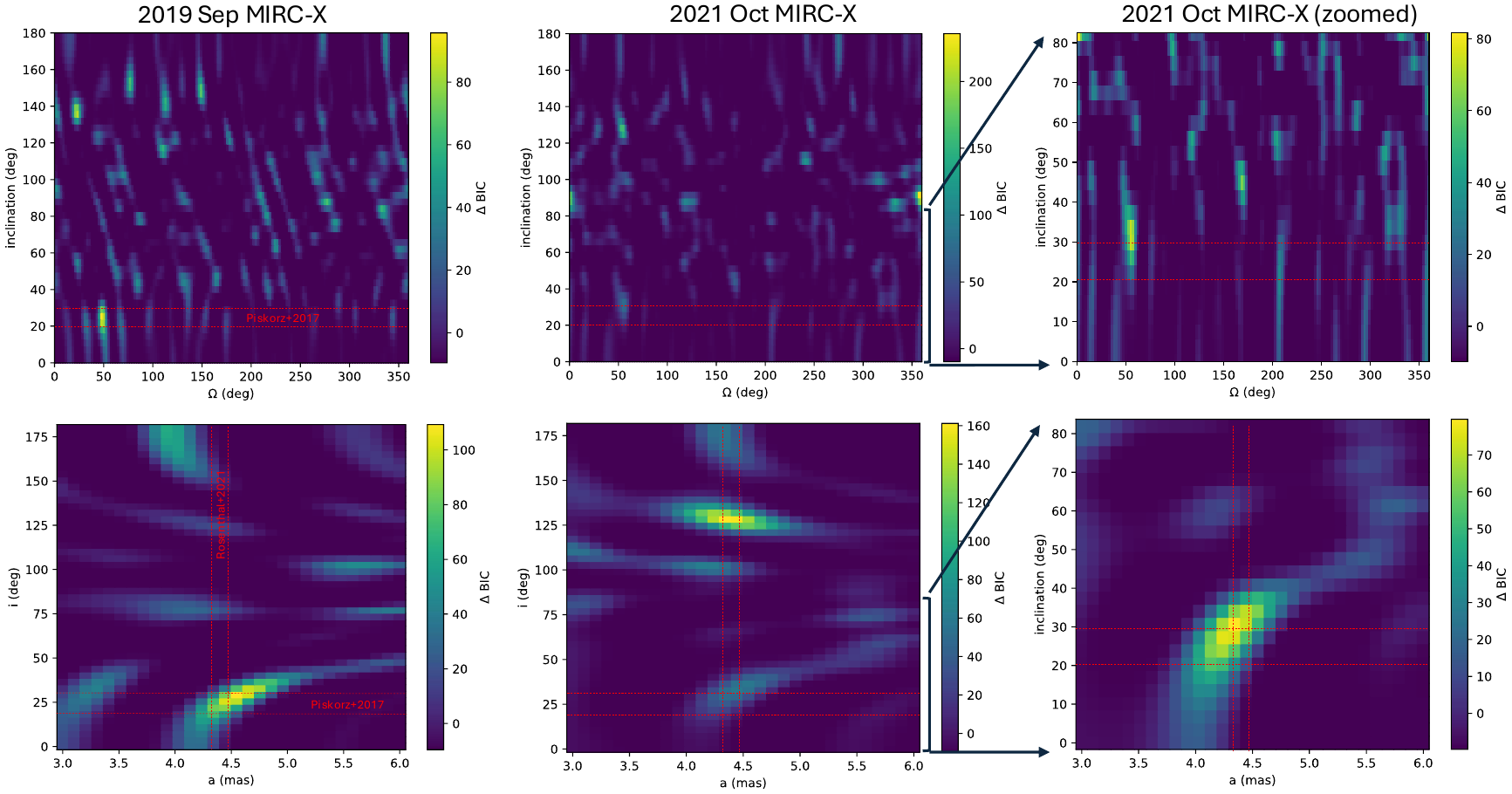}
\caption{ We show our grid searches for Ups And b to MIRC-X data across the unknown orbital elements from RV: $\Omega$, inclination, and semi-major axis. In 2019 September we find a $\chi^2$ minimum at $\Omega = 49^{\circ}$, $i=25^{\circ}$ (top left). We show the inclination/semi-major axis grid search at the best value of $\Omega$, and find $a=4.5$ mas (bottom left). The value for inclination agrees remarkably well with the results of \citet{piskorz2017}, and the semi-major axis is also consistent with previous measurements \citep[e.g.][]{rosenthal2021} (Top center) In the 2021 October MIRC-X dataset our best solution is an inclined orbit near 90$^{\circ}$, which we know is not real due to the lack of transits. However, we also see a solution near $\Omega = 53^{\circ}$, though we show that this solution prefers an inclination of $i=130^{\circ}$ (bottom center). If we consider only prograde orbits between 0--80$^{\circ}$, we find the same solution as the 2019 September dataset (right panels). As can be seen from the multiple solutions across different runs, this detection is extremely tentative and requires follow-up data to confirm. }
\label{upsand_tentative}
\end{figure}

Our first run was taken across 5 nights in September 2019. At the time, only the MIRC-X instrument was in commission. After running our data cleaning method, self-calibration, and grid search routine described in Section \ref{analysis} we found a potential solution at the same semi-major axis and inclination reported by \citet{piskorz2017}. The best-fit location of our semi-major axis is $a=4.4$ mas (in agreement with the distance and mass of the star), our inclination is at $i=25^{\circ}$, and $\Omega=49^{\circ}$. The planet/star contrast is 2.2$\times$10$^{-4}$, meaning that our instrument upgrades have improved our contrast levels from the previous MIRC upper limits reported in \citet{zhao2011}. This contrast is also broadly consistent with the model spectra reported in that work. Figure \ref{upsand_tentative} shows our tentative detection of the exoplanet. For the color map we choose to show the change in the Bayesian Information Criteria (BIC) between a uniform disk model (i.e. no planet) and a uniform disk + planet model. This is similar to the periodogram power used to search for planets in the ExoGRAVITY program, for example \citep{nowak2020}. The change in BIC is computed as

\begin{equation}
    \Delta BIC = (\chi_{UD}^2 + k_{UD} \ln n) - (\chi_{planet}^2 + k_{planet} \ln n),
\end{equation}

where $k$ is the number of free parameters and $n$ is the number of data points. The subscript ``UD" refers to a single uniform disk model, where ``planet" includes a planet companion. The model with the lower BIC is the better fit to the data, with $\Delta$BIC $>$ 10 typically representing a significantly better model \citep{liddle2007,Motalebi2015}. As can be seen in Figure \ref{upsand_tentative}, in our case there are multiple spots in the grid search map which represent significantly better fits to our data than the single star model (though this does not necessarily mean that these models are correct). 

Although this tentative detection is somewhat promising because of the agreement of parameters $a$ and $i$, we stress that the detection is marginal because there are many spots in the grid searches with similarly significant solutions. Note that we are fitting a monochromatic flux ratio at this time. When splitting the data into multiple spectral channels, our best detection location changes in these grid searches. Moreover, if we allow each wavelength channel to have its own fitted contrast the resulting H-band spectra is too noisy to reliably report. Hence, while this detection is certainly intriguing, we proposed for follow-up observations for confirmation and increased SNR. 

In 2021 October, we successfully observed the system across 4 nights with MIRC-X. After running our full pipeline of data cleaning and self-calibration, we search for the planet to see if it shares a detection with September 2019. The best solution is at an $\Omega$ near 360$^{\circ}$, and an edge-on inclination of nearly 90$^{\circ}$. Since there are no transits viewed for this system, we are confident this is not a correct solution. Searching near the 2019Sep detection, we see that there are two potential solutions which share a similar $\Omega$ value. However, the preferred solution is at an inclination of 130$^{\circ}$, which we do not see in 2019Sep. If we further restrict the grid searches to only include prograde orbits with inclination $<$80$^{\circ}$, we then recover a solution very similar to that of 2019. This solution has parameters of $a=4.4$mas, $i=23^{\circ}$, and $\Omega=53^{\circ}$ (see Figure \ref{upsand_tentative}). The monochromatic contrast ratio here is at a 3$\times$10$^{-4}$ level, though we again have insufficient SNR to break it into individual spectral channels. Given the multiple peaks in the grid search, it is obvious that these detections are still extremely tentative.

\begin{figure}[H]
\centering
\includegraphics[width=\textwidth]{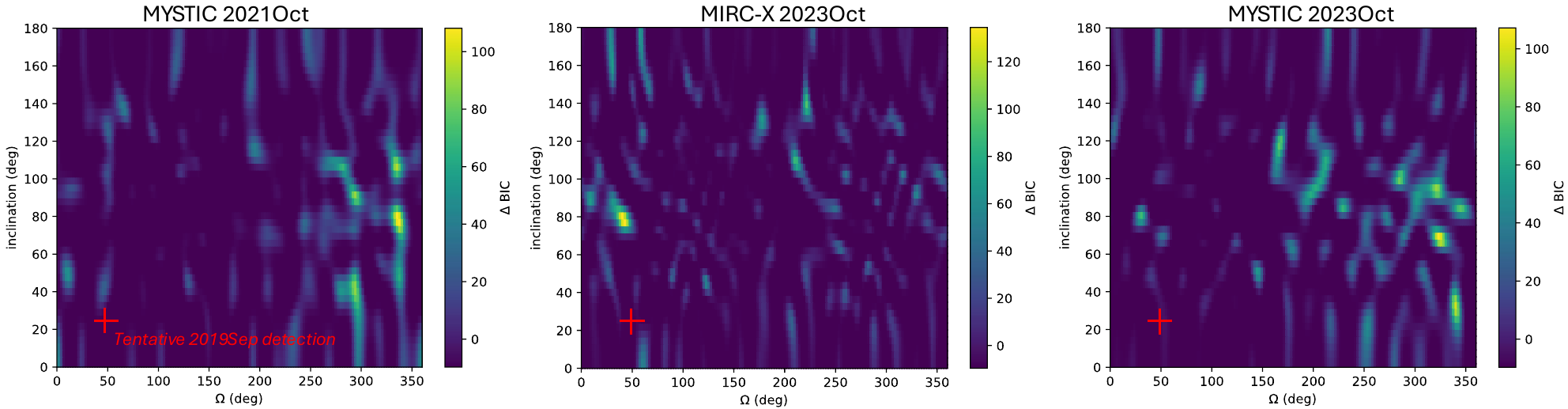}
\caption{ Although we had a very tentative detection of Ups And b in independent 2019Sep and 2021Oct MIRC-X datasets, we do not see this same detection in MYSTIC data from the same 2021Oct run (left). (Center and right) we also do not detect the planet at this location in a follow-up 2023Oct run with either MIRC-X or MYSTIC. }
\label{upsand_nondetection}
\end{figure}

As mentioned in previous sections, we took data with the K-band MYSTIC instrument simultaneously with MIRC-X during the 2021 October run. We also took a joint MIRC-X+MYSTIC dataset in 2023 October. This gives us additional independent datasets with which we might be able to confirm these tentative Ups And b detections. We do not find the planet at this location in the 2021 MYSTIC dataset, or in the MIRC-X or MYSTIC datasets of 2023 (Figure \ref{upsand_nondetection}). Moreover we do not see any similarly convincing shared peaks in the $\chi^2$ maps. This non-detection is puzzling, since including MYSTIC should help us given that the K-band contrast ratio should be more favorable than the H-band for a Hot Jupiter-type planet. Though our 2021 MYSTIC and 2023 MIRC-X+MYSTIC data quality did not seem any worse than previous sets, we first want to investigate whether or not we are probing deeper contrasts in the 2019+2021 MIRC-X datasets. In the following section, we present updated model spectra for this planet in the H and K band which we can then use for injection tests. 

\section{Updated Model Spectra and Injection Tests}\label{spectra}
\subsection{Ups And b Near IR Model Spectra}
In order to investigate the model spectra for Ups And b across the H and K bands, we first need updated global circulation models (GCMs) for the planet's atmosphere that can be turned into model spectra. \citet{malsky2021} recently computed GCMs for Ups And b, using the RM-GCM described in \citet{rauscher2012} and \citet{roman2017}. The values assumed for Ups And b matched those of recent high precision spectroscopy work of \citet{piskorz2017}, with a planet mass of 1.7 M$_{Jup}$. The orbital period was set to 4.617 days, semi-major axis 0.0595 au, orbital inclination of 24$^{\circ}$, stellar effective temperature of 6212 K, and stellar radius of 1.0296x10$^9$ meters (these choices are justified in \citealt{malsky2021}). These authors reported model spectra as a function of orbital phase in their paper, though the wavelength regime did not match our observations. We revised the calculations of \citet{malsky2021} to produce spectra in the H and K band, for both their clear and cloudy models of the Hot Jupiter atmosphere. As expected for a highly irradiated Hot Jupiter exoplanet, these updated spectra confirm that the flux contrast should be more favorable in the K-band. We find this to be the case for both clear and cloudy models for Ups And b. These model spectra certainly call into doubt the tentative MIRC-X detection, since we do not see the planet in K-band with our MYSTIC data taken at the same epoch. 

\begin{figure}[H]
\centering
\includegraphics[width=\textwidth]{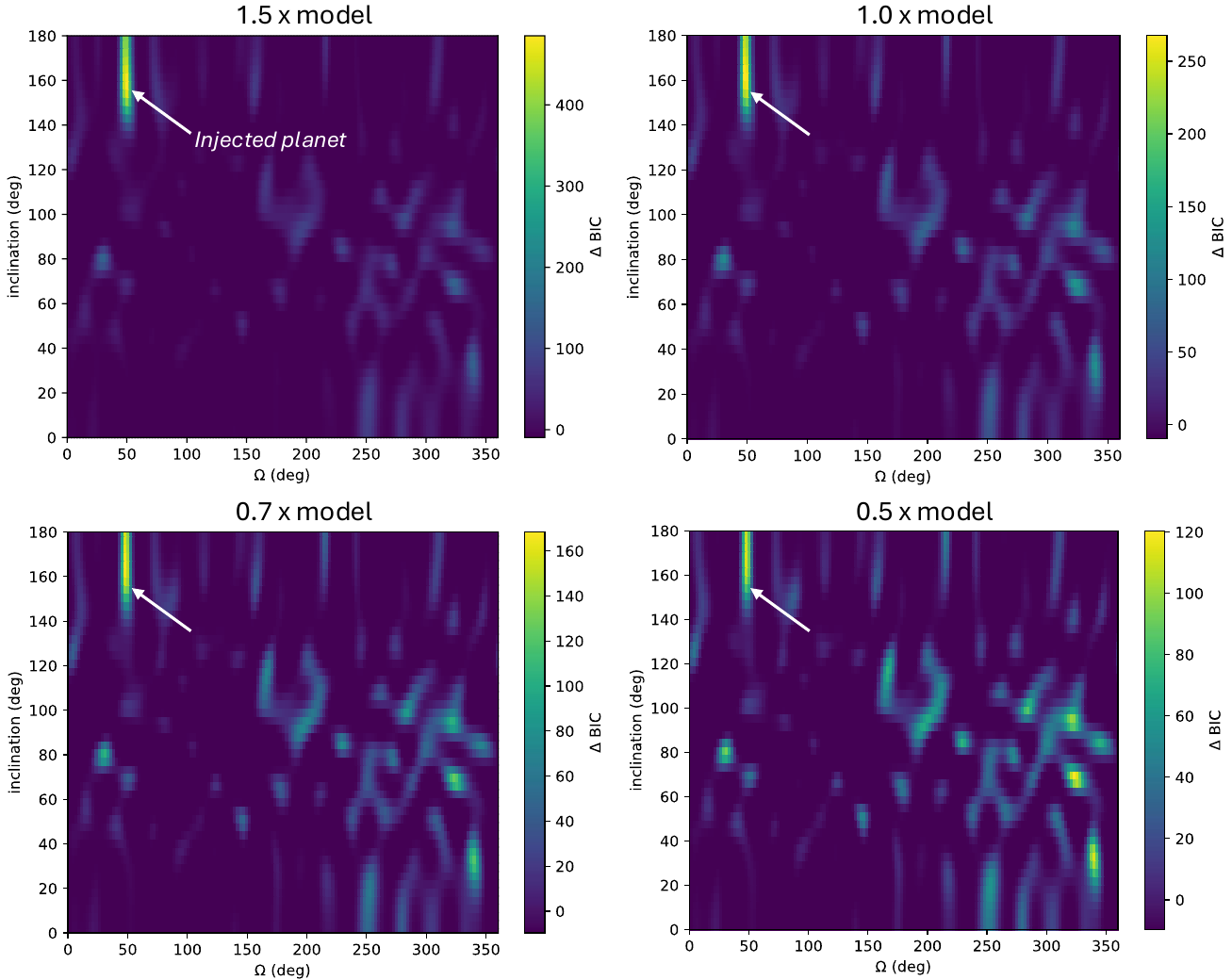}
\caption{ In order to test MYSTIC contrast limits, we inject planet signals of varying star/planet contrasts. Here we show a planet at an inclination of 156$^{\circ}$ and the same $\Omega$ as the MIRC-X detection. The planet signal is recovered down to a contrast of 0.7 times the value of the model planet flux. At fainter values than this, other peaks in the grid search become more prominent. This would suggest that, if the 2019/2021 MIRC-X detections were real, we should have recovered the planet again. }
\label{mystic_injection}
\end{figure}

\subsection{Injection Tests for Contrast Limits}

\begin{figure}[H]
\centering
\includegraphics[width=\textwidth]{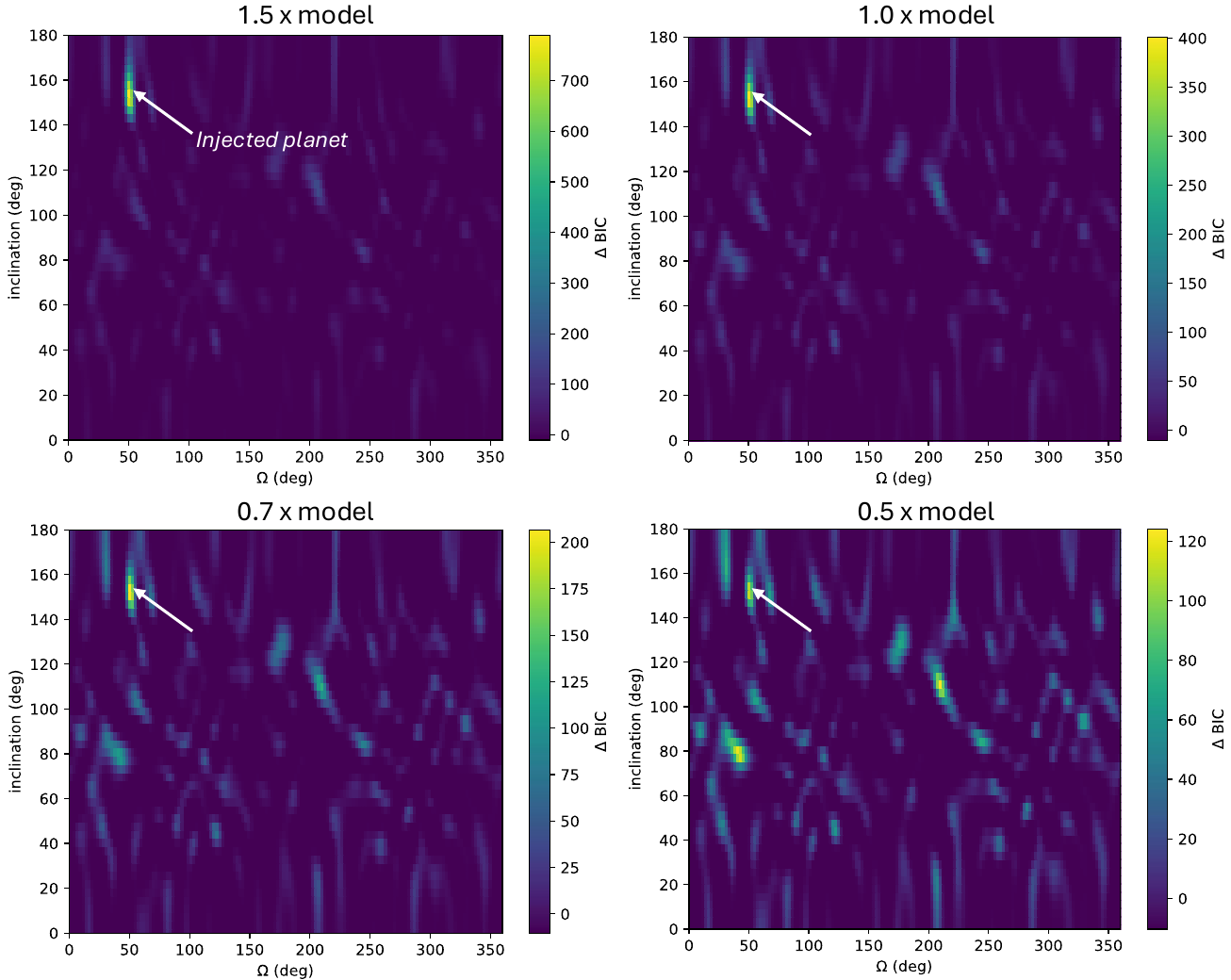}
\caption{ Same as Figure \ref{mystic_injection}, but for 2023Oct MIRC-X data. Just as for MYSTIC, we recover the planet down to a factor of 0.7 times the model flux from the planet. Once we get to 0.5 times the model flux and lower, other solutions become more prominent in the grid searches. }
\label{mircx_injection}
\end{figure}

To investigate the contrast limits, we inject a planet signal into the raw closure phase of our 2023 MIRC-X+MYSTIC data. This injected planet has the RV orbital elements shown in Section \ref{analysis}, and we added planets at four different locations of $\Omega$ / $i$ (51/24$^{\circ}$, 51/156$^{\circ}$, 231/24$^{\circ}$, 231/156$^{\circ}$, one for each quadrant of the grid search). We run injections for multiple different planet/star contrasts to test our limits. For the planet/star spectra, we use the models presented in the previous section and multiply them by factors from 0.5--1.5 in order to test different contrast levels. We then ran our data through the same cleaning and self-calibration pipelines as before, to see whether or not we could recover the signal from the injected planets. Figures \ref{mystic_injection} and \ref{mircx_injection} show the results of this test for the 2023Oct MYSTIC and MIRC-X datasets, respectively, at an injection location of $\Omega$ / $i$ = 51/156$^{\circ}$. As can be seen, the injected planet is recovered as the best solution down to 0.7$\times$ the model spectra for both instruments at this location, and it is still visible in the searches down to higher contrasts. In each quadrant for MIRC-X, we recover the planet best solution down to 0.7$\times$ the model planet flux, except for $\Omega$ / $i$ = 231/156$^{\circ}$ where we recover the planet down to 0.9$\times$ the model. For MYSTIC, we recover down to 0.7$\times$ for each location other than $\Omega$ / $i$ = 231/24$^{\circ}$, where it is again 0.9$\times$ the model. The successful detections were recovered at the correct injected locations, with better precision on $\Omega$ than on inclination as can be seen from the figures. Since inclination is measured as a projected separation, the precision on this quantity is limited by the resolution of the instrument. For these recovery tests, we are only fitting to monochromatic flux ratios. We generally recover a value that is near the median of the injected spectra across the bandpass. With future successful detections, we would measure the flux ratio as a function of wavelength. Our results suggest we are probing contrasts fainter than expected for Ups And b from the modeling described in the next section. These tests suggest that the 2019/2021 MIRC-X tentative detections are likely not real, since we should have recovered this same solution in 2021 MYSTIC and 2023 MIRC-X+MYSTIC.

\begin{figure}[H]
\centering
\includegraphics[width=\textwidth]{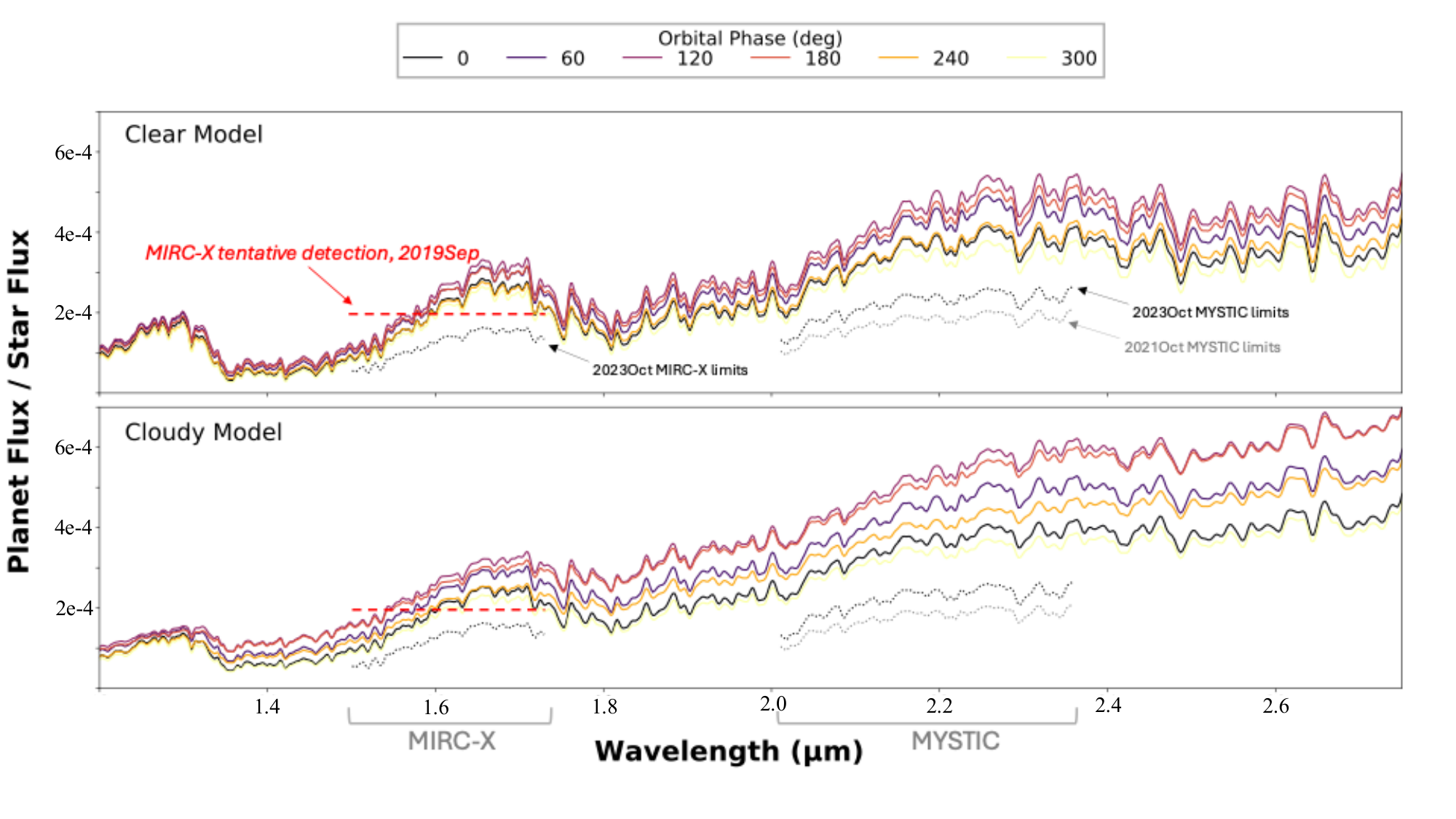}
\caption{ We use the GCMs of \citet{malsky2021} to produce model spectra dependent on orbital phase for Ups And b spanning the H and K bands (assuming an inclination of 24$^{\circ}$), for both a clear and cloudy atmosphere case (plotted as a planet-to-star flux ratio). As can be seen, our tentative MIRC-X monochromatic detection has a flux ratio which is broadly consistent with the models. As expected, this planet should have more favorable contrast in K-band. This is the wavelength where instruments such as MYSTIC and GRAVITY operate. We plot in the black dotted line the faintest recovered planet from our injection tests to the 2023Oct dataset. This shows that we can recover injected planets below the model values of flux ratio for both MIRC-X and MYSTIC, meaning that if the previous tentative detections were real, we should have recovered them again. }
\label{upsand_spectra}
\end{figure}

In Figure \ref{upsand_spectra} we show our model spectra in the near IR from updated GCMs of Ups And b, along with the limits we computed from injection tests. We plot the median of the limits at different $\Omega$ / $i$ for 2021 MYSTIC and 2023 MIRC-X+MYSTIC. In both cases, we are recovering injected planets lower than the model spectra and lower than the contrast of the tentative detections from 2019 and 2021 MIRC-X. Though we are clearly approaching the contrast levels to be able to detect Hot Jupiters with instruments like MIRC-X and MYSTIC, it is clear that we need an abundance of data across multiple runs to confirm tentative detections. A better understanding of our instrument model with regards to closure phase drifts due to polarization effects, dispersion, and jumps with altitude and azimuth is likely needed in order to most confidently calibrate our high precision closure phase data.

\section{Conclusions and Prospects for Future}\label{conclusions}
\begin{figure}[H]
\centering
\includegraphics[width=0.55\textwidth]{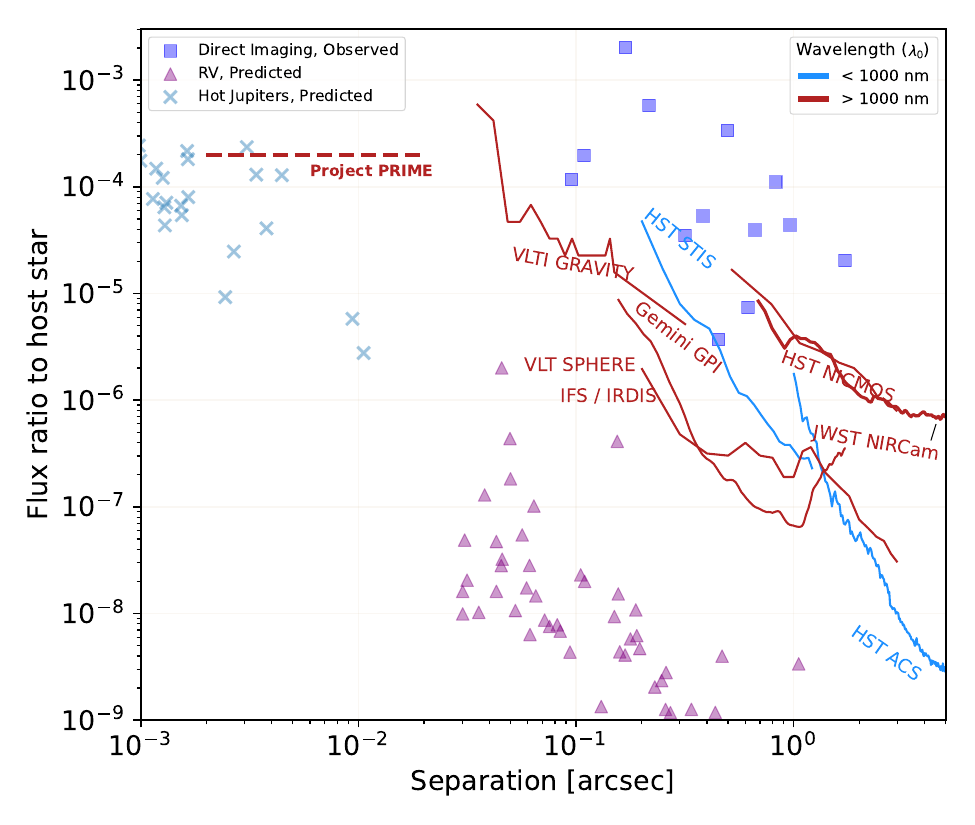}
\caption{ We include our Project PRIME contrast limits achieved in this work to the plot of \citet{pourre2024} (Fig. 12 in that paper, using the publicly available github repository at \url{https://github.com/nasavbailey/DI-flux-ratio-plot}). While the ExoGRAVITY team has used VLTI-GRAVITY to improve the inner working angle compared to single-telescope methods, we show that a program like PRIME has the potential to probe even closer to the host star by using the full resolution of long-baseline interferometers. We add estimated flux ratios of known hot Jupiter planets to this plot (using effective temperatures of the planet and star, along with the size ratio) to show that we are nearing the contrast regime for characterizing such systems.}
\label{flux_ratio_planets}
\end{figure}

Long baseline interferometry is an emerging tool for characterizing Hot Jupiters and other close-in high contrast systems. These systems are too close to their host star to be directly detected with single telescopes, but interferometers such at CHARA and VLTI are able to resolve the star--planet separation. One can then model the system as a high contrast ``binary", which reveals the visual orbital elements and planet/star flux ratio as a function of orbital phase. As demonstrated with our injection tests which can recover an exoplanet signal down to a contrast of 1--2$\times$10$^{-4}$ in H and K bands, we are at the cusp of being able to characterize the atmospheres of close-in giant planets with long baseline interferometry. However, our initially promising tentative MIRC-X detection is likely not real, since we are unable to confirm the planet location in the K-band with MYSTIC data or in a follow-up 2023 run with MIRC-X+MYSTIC. In Figure \ref{flux_ratio_planets} we show how our Project PRIME contrast limits compare to other methods, both interferometric and with single-telescopes. Techniques like PRIME could be instrumental in characterizing close-in hot companions in the near future.  

Currently, our closure phase precision is limited by instrument systematics. Hence, these detections require an intensive cleaning and self-calibration routine, which is crucial for achieving the sub-degree closure phase precision needed to approach these contrast levels. In this work we are calibrating closure phase drifts with a model dependent on altitude and azimuth, using the target data itself to calibrate such trends. However, even after this cleaning routine we are still dominated by systematics, preventing us from reaching the fundamental limits of the instruments.
Future work for the MIRC-X/MYSTIC team includes studying instrument polarization as a function of sky position, along with incorporating machine learning methods into our modeling of instrument systematics. In principle, one can build an instrument model to apply to the visibility and phase signals of all datasets taken with these instruments. If successful this could increase the closure phase accuracy and precision of MIRC-X and MYSTIC, allowing us to probe down to deeper contrasts. This future work will be a valuable addition to project PRIME. 

As shown in Figure \ref{upsand_spectra}, long baseline interferometry can add valuable datasets to Hot Jupiter atmosphere modeling. If a successful detection is made with enough SNR, these datasets can measure how the spectra changes as a function of orbital phase. The MIRC-X instrument has demonstrated the capability to reach deep contrasts in \citet{zhao2011} and with our tentative detection presented here, and K-band instruments such as GRAVITY and MYSTIC also have these capabilities along with more favorable contrasts for Hot Jupiters. Although there are not yet published results with the GRAVITY instrument, our team has received time to carry out similar studies with this instrument. Future work will include prospects for making similar types of exoplanet detection with VLTI/GRAVITY. Other future instruments such as NOTT (part of the planned ASGARD suite of visitor instruments at VLTI) are using nulling methods to improve contrast limits for close-in companions such as Hot Jupiters \citep{nott2023}. If these techniques can be perfected, long baseline interferometry will become a powerful tool for characterizing planets and other high contrast objects in the era of Gaia. The Gaia mission will discover thousands of exoplanets in astrometry, most of which will be too close to their host star to directly detect. Projects like ExoGRAVITY at VLTI are pushing to characterize planets down to $\sim$50 mas separations from their host star, and developing methods such as Project PRIME at CHARA have the potential to push such characterization techniques down to a few mas from the host star.

\section{Acknowledgements}
TG acknowledges support from from ERC Consolidator Grant (Grant Agreement ID 101003096) and Michigan Space Grant Consortium, NASA grant NNX15AJ20H. Some of the time at the CHARA Array was granted through the NOIRLab community access program (NOIRLab PropID: 2023B-519577; PI: T. Gardner).

SK acknowledges funding for MIRC-X received from the European Research Council (ERC) under the European Union's Horizon 2020 research and innovation programme (Starting Grant No. 639889 and Consolidated Grant No. 101003096) and STFC Consolidated Grant (ST/V000721/1) and STFC Small Award (ST/Y002695/1). JDM acknowledges funding for the development of MIRC-X (NASA-XRP NNX16AD43G, NSF-AST 2009489) and MYSTIC (NSF-ATI 1506540, NSF-AST 1909165).

This work is based upon observations obtained with the Georgia State University Center for High Angular Resolution Astronomy Array at Mount Wilson Observatory.  The CHARA Array is supported by the National Science Foundation under Grant No. AST-2034336 and AST-2407956.  Institutional support has been provided from the GSU College of Arts and Sciences and the GSU Office of the Vice President for Research and Economic Development.

A portion of this research was carried out at the Jet Propulsion Laboratory, California Institute of Technology, under a contract with the National Aeronautics and Space Administration (80NM0018D0004).

TG would like to thank Dr Shweta Dalal for years of useful comments and support.

The authors would like to thank the anonymous referee for helpful comments which improved the quality of this paper.

\section{Facilities and Software Used}
\textit{Facilities:} CHARA

\textit{Software:} Astropy \citep{astropy}, NumPy \citep{numpy}, SciPy \citep{scipy}


\bibliography{references}{}

\begin{thebibliography}{}
\expandafter\ifx\csname natexlab\endcsname\relax\def\natexlab#1{#1}\fi
\providecommand{\url}[1]{\href{#1}{#1}}
\providecommand{\dodoi}[1]{doi:~\href{http://doi.org/#1}{\nolinkurl{#1}}}
\providecommand{\doeprint}[1]{\href{http://ascl.net/#1}{\nolinkurl{http://ascl.net/#1}}}
\providecommand{\doarXiv}[1]{\href{https://arxiv.org/abs/#1}{\nolinkurl{https://arxiv.org/abs/#1}}}

\bibitem[{{Anugu} {et~al.}(2020){Anugu}, {Le Bouquin}, {Monnier}, {Kraus}, {Setterholm}, {Labdon}, {Davies}, {Lanthermann}, {Gardner}, {Ennis}, {Johnson}, {Ten Brummelaar}, {Schaefer}, \& {Sturmann}}]{anugu2020}
{Anugu}, N., {Le Bouquin}, J.-B., {Monnier}, J.~D., {et~al.} 2020, \aj, 160, 158, \dodoi{10.3847/1538-3881/aba957}

\bibitem[{{Astropy Collaboration} {et~al.}(2013){Astropy Collaboration}, {Robitaille}, {Tollerud}, {Greenfield}, {Droettboom}, {Bray}, {Aldcroft}, {Davis}, {Ginsburg}, {Price-Whelan}, {Kerzendorf}, {Conley}, {Crighton}, {Barbary}, {Muna}, {Ferguson}, {Grollier}, {Parikh}, {Nair}, {Unther}, {Deil}, {Woillez}, {Conseil}, {Kramer}, {Turner}, {Singer}, {Fox}, {Weaver}, {Zabalza}, {Edwards}, {Azalee Bostroem}, {Burke}, {Casey}, {Crawford}, {Dencheva}, {Ely}, {Jenness}, {Labrie}, {Lim}, {Pierfederici}, {Pontzen}, {Ptak}, {Refsdal}, {Servillat}, \& {Streicher}}]{astropy}
{Astropy Collaboration}, {Robitaille}, T.~P., {Tollerud}, E.~J., {et~al.} 2013, \aap, 558, A33, \dodoi{10.1051/0004-6361/201322068}

\bibitem[{{Balmer} {et~al.}(2024){Balmer}, {Franson}, {Chomez}, {Pueyo}, {Stolker}, {Lacour}, {Nowak}, {Nasedkin}, {Bonse}, {Thorngren}, {Palma-Bifani}, {Molliere}, {Wang}, {Zhang}, {Chavez}, {Kammerer}, {Blunt}, {Bowler}, {Bonnefoy}, {Brandner}, {Charnay}, {Chauvin}, {Henning}, {Lagrange}, {Pourre}, {Rickman}, {De Rosa}, {Vigan}, \& {Winterhalder}}]{balmer2024}
{Balmer}, W.~O., {Franson}, K., {Chomez}, A., {et~al.} 2024, arXiv e-prints, arXiv:2411.05917, \dodoi{10.48550/arXiv.2411.05917}

\bibitem[{{Basden} \& {Haniff}(2004)}]{basden2004}
{Basden}, A.~G., \& {Haniff}, C.~A. 2004, \mnras, 347, 1187, \dodoi{10.1111/j.1365-2966.2004.07283.x}

\bibitem[{{Brogi} {et~al.}(2013){Brogi}, {Snellen}, {de Kok}, {Albrecht}, {Birkby}, \& {de Mooij}}]{brogi2013}
{Brogi}, M., {Snellen}, I.~A.~G., {de Kok}, R.~J., {et~al.} 2013, \apj, 767, 27, \dodoi{10.1088/0004-637X/767/1/27}

\bibitem[{{Buzard} {et~al.}(2021){Buzard}, {Piskorz}, {Lockwood}, {Blake}, {Barman}, {Benneke}, {Bender}, \& {Carr}}]{buzard2021}
{Buzard}, C., {Piskorz}, D., {Lockwood}, A.~C., {et~al.} 2021, \aj, 162, 269, \dodoi{10.3847/1538-3881/ac2a2c}

\bibitem[{{Cowan} {et~al.}(2012){Cowan}, {Voigt}, \& {Abbot}}]{cowan2012}
{Cowan}, N.~B., {Voigt}, A., \& {Abbot}, D.~S. 2012, \apj, 757, 80, \dodoi{10.1088/0004-637X/757/1/80}

\bibitem[{{Fortney} {et~al.}(2021){Fortney}, {Dawson}, \& {Komacek}}]{fortney2021}
{Fortney}, J.~J., {Dawson}, R.~I., \& {Komacek}, T.~D. 2021, Journal of Geophysical Research (Planets), 126, e06629, \dodoi{10.1029/2020JE006629}

\bibitem[{{Gaia Collaboration} {et~al.}(2021){Gaia Collaboration}, {Brown}, {Vallenari}, {Prusti}, {de Bruijne}, {Babusiaux}, {Biermann}, {Creevey}, {Evans}, {Eyer}, {Hutton}, {Jansen}, {Jordi}, {Klioner}, {Lammers}, {Lindegren}, {Luri}, {Mignard}, {Panem}, {Pourbaix}, {Randich}, {Sartoretti}, {Soubiran}, {Walton}, {Arenou}, {Bailer-Jones}, {Bastian}, {Cropper}, {Drimmel}, {Katz}, {Lattanzi}, {van Leeuwen}, {Bakker}, {Cacciari}, {Casta{\~n}eda}, {De Angeli}, {Ducourant}, {Fabricius}, {Fouesneau}, {Fr{\'e}mat}, {Guerra}, {Guerrier}, {Guiraud}, {Jean-Antoine Piccolo}, {Masana}, {Messineo}, {Mowlavi}, {Nicolas}, {Nienartowicz}, {Pailler}, {Panuzzo}, {Riclet}, {Roux}, {Seabroke}, {Sordo}, {Tanga}, {Th{\'e}venin}, {Gracia-Abril}, {Portell}, {Teyssier}, {Altmann}, {Andrae}, {Bellas-Velidis}, {Benson}, {Berthier}, {Blomme}, {Brugaletta}, {Burgess}, {Busso}, {Carry}, {Cellino}, {Cheek}, {Clementini}, {Damerdji}, {Davidson}, {Delchambre}, {Dell'Oro}, {Fern{\'a}ndez-Hern{\'a}ndez}, {Galluccio}, {Garc{\'\i}a-Lario},
  {Garcia-Reinaldos}, {Gonz{\'a}lez-N{\'u}{\~n}ez}, {Gosset}, {Haigron}, {Halbwachs}, {Hambly}, {Harrison}, {Hatzidimitriou}, {Heiter}, {Hern{\'a}ndez}, {Hestroffer}, {Hodgkin}, {Holl}, {Jan{\ss}en}, {Jevardat de Fombelle}, {Jordan}, {Krone-Martins}, {Lanzafame}, {L{\"o}ffler}, {Lorca}, {Manteiga}, {Marchal}, {Marrese}, {Moitinho}, {Mora}, {Muinonen}, {Osborne}, {Pancino}, {Pauwels}, {Petit}, {Recio-Blanco}, {Richards}, {Riello}, {Rimoldini}, {Robin}, {Roegiers}, {Rybizki}, {Sarro}, {Siopis}, {Smith}, {Sozzetti}, {Ulla}, {Utrilla}, {van Leeuwen}, {van Reeven}, {Abbas}, {Abreu Aramburu}, {Accart}, {Aerts}, {Aguado}, {Ajaj}, {Altavilla}, {{\'A}lvarez}, {{\'A}lvarez Cid-Fuentes}, {Alves}, {Anderson}, {Anglada Varela}, {Antoja}, {Audard}, {Baines}, {Baker}, {Balaguer-N{\'u}{\~n}ez}, {Balbinot}, {Balog}, {Barache}, {Barbato}, {Barros}, {Barstow}, {Bartolom{\'e}}, {Bassilana}, {Bauchet}, {Baudesson-Stella}, {Becciani}, {Bellazzini}, {Bernet}, {Bertone}, {Bianchi}, {Blanco-Cuaresma}, {Boch}, {Bombrun}, {Bossini},
  {Bouquillon}, {Bragaglia}, {Bramante}, {Breedt}, {Bressan}, {Brouillet}, {Bucciarelli}, {Burlacu}, {Busonero}, {Butkevich}, {Buzzi}, {Caffau}, {Cancelliere}, {C{\'a}novas}, {Cantat-Gaudin}, {Carballo}, {Carlucci}, {Carnerero}, {Carrasco}, {Casamiquela}, {Castellani}, {Castro-Ginard}, {Castro Sampol}, {Chaoul}, {Charlot}, {Chemin}, {Chiavassa}, {Cioni}, {Comoretto}, {Cooper}, {Cornez}, {Cowell}, {Crifo}, {Crosta}, {Crowley}, {Dafonte}, {Dapergolas}, {David}, {David}, {de Laverny}, {De Luise}, {De March}, {De Ridder}, {de Souza}, {de Teodoro}, {de Torres}, {del Peloso}, {del Pozo}, {Delbo}, {Delgado}, {Delgado}, {Delisle}, {Di Matteo}, {Diakite}, {Diener}, {Distefano}, {Dolding}, {Eappachen}, {Edvardsson}, {Enke}, {Esquej}, {Fabre}, {Fabrizio}, {Faigler}, {Fedorets}, {Fernique}, {Fienga}, {Figueras}, {Fouron}, {Fragkoudi}, {Fraile}, {Franke}, {Gai}, {Garabato}, {Garcia-Gutierrez}, {Garc{\'\i}a-Torres}, {Garofalo}, {Gavras}, {Gerlach}, {Geyer}, {Giacobbe}, {Gilmore}, {Girona}, {Giuffrida}, {Gomel}, {Gomez},
  {Gonzalez-Santamaria}, {Gonz{\'a}lez-Vidal}, {Granvik}, {Guti{\'e}rrez-S{\'a}nchez}, {Guy}, {Hauser}, {Haywood}, {Helmi}, {Hidalgo}, {Hilger}, {H{\l}adczuk}, {Hobbs}, {Holland}, {Huckle}, {Jasniewicz}, {Jonker}, {Juaristi Campillo}, {Julbe}, {Karbevska}, {Kervella}, {Khanna}, {Kochoska}, {Kontizas}, {Kordopatis}, {Korn}, {Kostrzewa-Rutkowska}, {Kruszy{\'n}ska}, {Lambert}, {Lanza}, {Lasne}, {Le Campion}, {Le Fustec}, {Lebreton}, {Lebzelter}, {Leccia}, {Leclerc}, {Lecoeur-Taibi}, {Liao}, {Licata}, {Lindstr{\o}m}, {Lister}, {Livanou}, {Lobel}, {Madrero Pardo}, {Managau}, {Mann}, {Marchant}, {Marconi}, {Marcos Santos}, {Marinoni}, {Marocco}, {Marshall}, {Martin Polo}, {Mart{\'\i}n-Fleitas}, {Masip}, {Massari}, {Mastrobuono-Battisti}, {Mazeh}, {McMillan}, {Messina}, {Michalik}, {Millar}, {Mints}, {Molina}, {Molinaro}, {Moln{\'a}r}, {Montegriffo}, {Mor}, {Morbidelli}, {Morel}, {Morris}, {Mulone}, {Munoz}, {Muraveva}, {Murphy}, {Musella}, {Noval}, {Ord{\'e}novic}, {Orr{\`u}}, {Osinde}, {Pagani}, {Pagano},
  {Palaversa}, {Palicio}, {Panahi}, {Pawlak}, {Pe{\~n}alosa Esteller}, {Penttil{\"a}}, {Piersimoni}, {Pineau}, {Plachy}, {Plum}, {Poggio}, {Poretti}, {Poujoulet}, {Pr{\v{s}}a}, {Pulone}, {Racero}, {Ragaini}, {Rainer}, {Raiteri}, {Rambaux}, {Ramos}, {Ramos-Lerate}, {Re Fiorentin}, {Regibo}, {Reyl{\'e}}, {Ripepi}, {Riva}, {Rixon}, {Robichon}, {Robin}, {Roelens}, {Rohrbasser}, {Romero-G{\'o}mez}, {Rowell}, {Royer}, {Rybicki}, {Sadowski}, {Sagrist{\`a} Sell{\'e}s}, {Sahlmann}, {Salgado}, {Salguero}, {Samaras}, {Sanchez Gimenez}, {Sanna}, {Santove{\~n}a}, {Sarasso}, {Schultheis}, {Sciacca}, {Segol}, {Segovia}, {S{\'e}gransan}, {Semeux}, {Shahaf}, {Siddiqui}, {Siebert}, {Siltala}, {Slezak}, {Smart}, {Solano}, {Solitro}, {Souami}, {Souchay}, {Spagna}, {Spoto}, {Steele}, {Steidelm{\"u}ller}, {Stephenson}, {S{\"u}veges}, {Szabados}, {Szegedi-Elek}, {Taris}, {Tauran}, {Taylor}, {Teixeira}, {Thuillot}, {Tonello}, {Torra}, {Torra}, {Turon}, {Unger}, {Vaillant}, {van Dillen}, {Vanel}, {Vecchiato}, {Viala}, {Vicente},
  {Voutsinas}, {Weiler}, {Wevers}, {Wyrzykowski}, {Yoldas}, {Yvard}, {Zhao}, {Zorec}, {Zucker}, {Zurbach}, \& {Zwitter}}]{gaia_dr3}
{Gaia Collaboration}, {Brown}, A.~G.~A., {Vallenari}, A., {et~al.} 2021, \aap, 649, A1, \dodoi{10.1051/0004-6361/202039657}

\bibitem[{{Garcia} {et~al.}(2016){Garcia}, {Muterspaugh}, {van Belle}, {Monnier}, {Stassun}, {Ghasempour}, {Clark}, {Zavala}, {Benson}, {Hutter}, {Schmitt}, {Baines}, {Jorgensen}, {Strosahl}, {Sanborn}, {Zawicki}, {Sakosky}, \& {Swihart}}]{garcia2016}
{Garcia}, E.~V., {Muterspaugh}, M.~W., {van Belle}, G., {et~al.} 2016, \pasp, 128, 055004, \dodoi{10.1088/1538-3873/128/963/055004}

\bibitem[{Harris {et~al.}(2020)Harris, Millman, van~der Walt, Gommers, Virtanen, Cournapeau, Wieser, Taylor, Berg, Smith, Kern, Picus, Hoyer, van Kerkwijk, Brett, Haldane, del R{'{\i}}o, Wiebe, Peterson, G{'{e}}rard-Marchant, Sheppard, Reddy, Weckesser, Abbasi, Gohlke, \& Oliphant}]{numpy}
Harris, C.~R., Millman, K.~J., van~der Walt, S.~J., {et~al.} 2020, Nature, 585, 357, \dodoi{10.1038/s41586-020-2649-2}

\bibitem[{{Kreidberg} {et~al.}(2014){Kreidberg}, {Bean}, {D{\'e}sert}, {Line}, {Fortney}, {Madhusudhan}, {Stevenson}, {Showman}, {Charbonneau}, {McCullough}, {Seager}, {Burrows}, {Henry}, {Williamson}, {Kataria}, \& {Homeier}}]{kreidberg2014}
{Kreidberg}, L., {Bean}, J.~L., {D{\'e}sert}, J.-M., {et~al.} 2014, \apjl, 793, L27, \dodoi{10.1088/2041-8205/793/2/L27}

\bibitem[{Lacour(2023)}]{lacour2023}
Lacour, S. 2023, Comptes Rendus. Physique, 24, 115, \dodoi{10.5802/crphys.144}

\bibitem[{{Laugier} {et~al.}(2023){Laugier}, {Defr{\`e}re}, {Woillez}, {Courtney-Barrer}, {Dannert}, {Matter}, {Dandumont}, {Gross}, {Absil}, {Bigioli}, {Garreau}, {Labadie}, {Loicq}, {Martinod}, {Mazzoli}, {Raskin}, \& {Sanny}}]{nott2023}
{Laugier}, R., {Defr{\`e}re}, D., {Woillez}, J., {et~al.} 2023, \aap, 671, A110, \dodoi{10.1051/0004-6361/202244351}

\bibitem[{{Liddle}(2007)}]{liddle2007}
{Liddle}, A.~R. 2007, \mnras, 377, L74, \dodoi{10.1111/j.1745-3933.2007.00306.x}

\bibitem[{{Malsky} {et~al.}(2021){Malsky}, {Rauscher}, {Kempton}, {Roman}, {Long}, \& {Harada}}]{malsky2021}
{Malsky}, I., {Rauscher}, E., {Kempton}, E. M.~R., {et~al.} 2021, \apj, 923, 62, \dodoi{10.3847/1538-4357/ac2a2a}

\bibitem[{{May} {et~al.}(2018){May}, {Zhao}, {Haidar}, {Rauscher}, \& {Monnier}}]{may2018}
{May}, E.~M., {Zhao}, M., {Haidar}, M., {Rauscher}, E., \& {Monnier}, J.~D. 2018, \aj, 156, 122, \dodoi{10.3847/1538-3881/aad4a8}

\bibitem[{{McArthur} {et~al.}(2010){McArthur}, {Benedict}, {Barnes}, {Martioli}, {Korzennik}, {Nelan}, \& {Butler}}]{mcarthur2010}
{McArthur}, B.~E., {Benedict}, G.~F., {Barnes}, R., {et~al.} 2010, \apj, 715, 1203, \dodoi{10.1088/0004-637X/715/2/1203}

\bibitem[{{Monnier} {et~al.}(2006){Monnier}, {Pedretti}, {Thureau}, {Berger}, {Millan-Gabet}, {ten Brummelaar}, {McAlister}, {Sturmann}, {Sturmann}, {Muirhead}, {Tannirkulam}, {Webster}, \& {Zhao}}]{Monnier2006}
{Monnier}, J.~D., {Pedretti}, E., {Thureau}, N., {et~al.} 2006, in \procspie, Vol. 6268, Society of Photo-Optical Instrumentation Engineers (SPIE) Conference Series, 62681P, \dodoi{10.1117/12.671982}

\bibitem[{{Monnier} {et~al.}(2012){Monnier}, {Che}, {Zhao}, {Ekstr{\"o}m}, {Maestro}, {Aufdenberg}, {Baron}, {Georgy}, {Kraus}, {McAlister}, {Pedretti}, {Ridgway}, {Sturmann}, {Sturmann}, {ten Brummelaar}, {Thureau}, {Turner}, \& {Tuthill}}]{monnier2012}
{Monnier}, J.~D., {Che}, X., {Zhao}, M., {et~al.} 2012, \apjl, 761, L3, \dodoi{10.1088/2041-8205/761/1/L3}

\bibitem[{{Motalebi} {et~al.}(2015){Motalebi}, {Udry}, {Gillon}, {Lovis}, {S{\'e}gransan}, {Buchhave}, {Demory}, {Malavolta}, {Dressing}, {Sasselov}, {Rice}, {Charbonneau}, {Collier Cameron}, {Latham}, {Molinari}, {Pepe}, {Affer}, {Bonomo}, {Cosentino}, {Dumusque}, {Figueira}, {Fiorenzano}, {Gettel}, {Harutyunyan}, {Haywood}, {Johnson}, {Lopez}, {Lopez-Morales}, {Mayor}, {Micela}, {Mortier}, {Nascimbeni}, {Philips}, {Piotto}, {Pollacco}, {Queloz}, {Sozzetti}, {Vanderburg}, \& {Watson}}]{Motalebi2015}
{Motalebi}, F., {Udry}, S., {Gillon}, M., {et~al.} 2015, \aap, 584, A72, \dodoi{10.1051/0004-6361/201526822}

\bibitem[{{Nasedkin} {et~al.}(2024){Nasedkin}, {Molli{\`e}re}, {Lacour}, {Nowak}, {Kreidberg}, {Stolker}, {Wang}, {Balmer}, {Kammerer}, {Shangguan}, {Abuter}, {Amorim}, {Asensio-Torres}, {Benisty}, {Berger}, {Beust}, {Blunt}, {Boccaletti}, {Bonnefoy}, {Bonnet}, {Bordoni}, {Bourdarot}, {Brandner}, {Cantalloube}, {Caselli}, {Charnay}, {Chauvin}, {Chavez}, {Choquet}, {Christiaens}, {Cl{\'e}net}, {Coud{\'e} Du Foresto}, {Cridland}, {Davies}, {Dembet}, {Dexter}, {Drescher}, {Duvert}, {Eckart}, {Eisenhauer}, {F{\"o}rster Schreiber}, {Garcia}, {Garcia Lopez}, {Gendron}, {Genzel}, {Gillessen}, {Girard}, {Grant}, {Haubois}, {Hei{\ss}el}, {Henning}, {Hinkley}, {Hippler}, {Houll{\'e}}, {Hubert}, {Jocou}, {Keppler}, {Kervella}, {Kurtovic}, {Lagrange}, {Lapeyr{\`e}re}, {Le Bouquin}, {Lutz}, {Maire}, {Mang}, {Marleau}, {M{\'e}rand}, {Monnier}, {Mordasini}, {Ott}, {Otten}, {Paladini}, {Paumard}, {Perraut}, {Perrin}, {Pfuhl}, {Pourr{\'e}}, {Pueyo}, {Ribeiro}, {Rickman}, {Ruffio}, {Rustamkulov}, {Shimizu}, {Sing}, {Stadler},
  {Straub}, {Straubmeier}, {Sturm}, {Tacconi}, {van Dishoeck}, {Vigan}, {Vincent}, {von Fellenberg}, {Widmann}, {Winterhalder}, {Woillez}, {Yazici}, \& {Gravity Collaboration}}]{nasedkin2024}
{Nasedkin}, E., {Molli{\`e}re}, P., {Lacour}, S., {et~al.} 2024, \aap, 687, A298, \dodoi{10.1051/0004-6361/202449328}

\bibitem[{{Newville} {et~al.}(2016){Newville}, {Stensitzki}, {Allen}, {Rawlik}, {Ingargiola}, \& {Nelson}}]{newville2016}
{Newville}, M., {Stensitzki}, T., {Allen}, D.~B., {et~al.} 2016, {Lmfit: Non-Linear Least-Square Minimization and Curve-Fitting for Python}.
\newblock \doeprint{1606.014}

\bibitem[{{Nikolov} {et~al.}(2014){Nikolov}, {Sing}, {Pont}, {Burrows}, {Fortney}, {Ballester}, {Evans}, {Huitson}, {Wakeford}, {Wilson}, {Aigrain}, {Deming}, {Gibson}, {Henry}, {Knutson}, {Lecavelier des Etangs}, {Showman}, {Vidal-Madjar}, \& {Zahnle}}]{nikolov2014}
{Nikolov}, N., {Sing}, D.~K., {Pont}, F., {et~al.} 2014, \mnras, 437, 46, \dodoi{10.1093/mnras/stt1859}

\bibitem[{{Nowak} {et~al.}(2020){Nowak}, {Lacour}, {Lagrange}, {Rubini}, {Wang}, {Stolker}, {Abuter}, {Amorim}, {Asensio-Torres}, {Baub{\"o}ck}, {Benisty}, {Berger}, {Beust}, {Blunt}, {Boccaletti}, {Bonnefoy}, {Bonnet}, {Brandner}, {Cantalloube}, {Charnay}, {Choquet}, {Christiaens}, {Cl{\'e}net}, {Coud{\'e} Du Foresto}, {Cridland}, {de Zeeuw}, {Dembet}, {Dexter}, {Drescher}, {Duvert}, {Eckart}, {Eisenhauer}, {Gao}, {Garcia}, {Garcia Lopez}, {Gardner}, {Gendron}, {Genzel}, {Gillessen}, {Girard}, {Grandjean}, {Haubois}, {Hei{\ss}el}, {Henning}, {Hinkley}, {Hippler}, {Horrobin}, {Houll{\'e}}, {Hubert}, {Jim{\'e}nez-Rosales}, {Jocou}, {Kammerer}, {Kervella}, {Keppler}, {Kreidberg}, {Kulikauskas}, {Lapeyr{\`e}re}, {Le Bouquin}, {L{\'e}na}, {M{\'e}rand}, {Maire}, {Molli{\`e}re}, {Monnier}, {Mouillet}, {M{\"u}ller}, {Nasedkin}, {Ott}, {Otten}, {Paumard}, {Paladini}, {Perraut}, {Perrin}, {Pueyo}, {Pfuhl}, {Rameau}, {Rodet}, {Rodr{\'\i}guez-Coira}, {Rousset}, {Scheithauer}, {Shangguan}, {Stadler}, {Straub},
  {Straubmeier}, {Sturm}, {Tacconi}, {van Dishoeck}, {Vigan}, {Vincent}, {von Fellenberg}, {Ward-Duong}, {Widmann}, {Wieprecht}, {Wiezorrek}, {Woillez}, \& {Gravity Collaboration}}]{nowak2020}
{Nowak}, M., {Lacour}, S., {Lagrange}, A.~M., {et~al.} 2020, \aap, 642, L2, \dodoi{10.1051/0004-6361/202039039}

\bibitem[{{Parmentier} \& {Crossfield}(2018)}]{parmentier2018}
{Parmentier}, V., \& {Crossfield}, I. J.~M. 2018, in Handbook of Exoplanets, ed. H.~J. {Deeg} \& J.~A. {Belmonte}, 116, \dodoi{10.1007/978-3-319-55333-7_116}

\bibitem[{{Piskorz} {et~al.}(2017){Piskorz}, {Benneke}, {Crockett}, {Lockwood}, {Blake}, {Barman}, {Bender}, {Carr}, \& {Johnson}}]{piskorz2017}
{Piskorz}, D., {Benneke}, B., {Crockett}, N.~R., {et~al.} 2017, \aj, 154, 78, \dodoi{10.3847/1538-3881/aa7dd8}

\bibitem[{{Pourr{\'e}} {et~al.}(2024){Pourr{\'e}}, {Winterhalder}, {Le Bouquin}, {Lacour}, {Bidot}, {Nowak}, {Maire}, {Mouillet}, {Babusiaux}, {Woillez}, {Abuter}, {Amorim}, {Asensio-Torres}, {Balmer}, {Benisty}, {Berger}, {Beust}, {Blunt}, {Boccaletti}, {Bonnefoy}, {Bonnet}, {Bordoni}, {Bourdarot}, {Brandner}, {Cantalloube}, {Caselli}, {Charnay}, {Chauvin}, {Chavez}, {Choquet}, {Christiaens}, {Cl{\'e}net}, {du Foresto}, {Cridland}, {Davies}, {Defr{\`e}re}, {Dembet}, {Dexter}, {Drescher}, {Duvert}, {Eckart}, {Eisenhauer}, {Schreiber}, {Garcia}, {Lopez}, {Gendron}, {Genzel}, {Gillessen}, {Girard}, {Gonte}, {Grant}, {Haubois}, {Hei{\ss}el}, {Henning}, {Hinkley}, {Hippler}, {H{\"o}nig}, {Houll{\'e}}, {Hubert}, {Jocou}, {Kammerer}, {Kenworthy}, {Keppler}, {Kervella}, {Kreidberg}, {Kurtovic}, {Lagrange}, {Lapeyr{\`e}re}, {Lutz}, {Mang}, {Marleau}, {M{\'e}rand}, {Millour}, {Molli{\`e}re}, {Monnier}, {Mordasini}, {Nasedkin}, {Oberti}, {Ott}, {Otten}, {Paladini}, {Paumard}, {Perraut}, {Perrin}, {Pfuhl}, {Pueyo},
  {Ribeiro}, {Rickman}, {Rustamkulov}, {Shangguan}, {Shimizu}, {Sing}, {Soulez}, {Stadler}, {Stolker}, {Straub}, {Straubmeier}, {Sturm}, {Sykes}, {Tacconi}, {van Dishoeck}, {Vigan}, {Vincent}, {von Fellenberg}, {Wang}, {Widmann}, {Yazici}, {Abad}, {Carpentier}, {Alonso}, {Andolfato}, {Barriga}, {Beuzit}, {Bourget}, {Brast}, {Caniguante}, {Cottalorda}, {Darr{\'e}}, {Delabre}, {Delboulb{\'e}}, {Delplancke-Str{\"o}bele}, {Donaldson}, {Dorn}, {Dupuy}, {Egner}, {Fischer}, {Frank}, {Fuenteseca}, {Gitton}, {Guerlet}, {Guieu}, {Gutierrez}, {Haguenauer}, {Haimerl}, {Heritier}, {Huber}, {Hubin}, {Jolley}, {Kirchbauer}, {Kolb}, {Kosmalski}, {Krempl}, {Le Louarn}, {Lilley}, {Lopez}, {Magnard}, {Mclay}, {Meilland}, {Meister}, {Moulin}, {Pasquini}, {Paufique}, {Percheron}, {Pettazzi}, {Phan}, {Pirani}, {Quentin}, {Rakich}, {Ridings}, {Reyes}, {Rochat}, {Schmid}, {Schuhler}, {Shchekaturov}, {Seidel}, {Soenke}, {Stadler}, {Stephan}, {Su{\'a}rez}, {Todorovic}, {Valdes}, {Verinaud}, {Zins}, \&
  {Z{\'u}{\~n}iga-Fern{\'a}ndez}}]{pourre2024}
{Pourr{\'e}}, N., {Winterhalder}, T.~O., {Le Bouquin}, J.~B., {et~al.} 2024, \aap, 686, A258, \dodoi{10.1051/0004-6361/202449507}

\bibitem[{{Rauscher} \& {Menou}(2012)}]{rauscher2012}
{Rauscher}, E., \& {Menou}, K. 2012, \apj, 750, 96, \dodoi{10.1088/0004-637X/750/2/96}

\bibitem[{{Rodler} {et~al.}(2012){Rodler}, {Lopez-Morales}, \& {Ribas}}]{rodler2012}
{Rodler}, F., {Lopez-Morales}, M., \& {Ribas}, I. 2012, \apjl, 753, L25, \dodoi{10.1088/2041-8205/753/1/L25}

\bibitem[{{Roman} \& {Rauscher}(2017)}]{roman2017}
{Roman}, M., \& {Rauscher}, E. 2017, \apj, 850, 17, \dodoi{10.3847/1538-4357/aa8ee4}

\bibitem[{{Rosenthal} {et~al.}(2021){Rosenthal}, {Fulton}, {Hirsch}, {Isaacson}, {Howard}, {Dedrick}, {Sherstyuk}, {Blunt}, {Petigura}, {Knutson}, {Behmard}, {Chontos}, {Crepp}, {Crossfield}, {Dalba}, {Fischer}, {Henry}, {Kane}, {Kosiarek}, {Marcy}, {Rubenzahl}, {Weiss}, \& {Wright}}]{rosenthal2021}
{Rosenthal}, L.~J., {Fulton}, B.~J., {Hirsch}, L.~A., {et~al.} 2021, \apjs, 255, 8, \dodoi{10.3847/1538-4365/abe23c}

\bibitem[{{Samra} {et~al.}(2023){Samra}, {Helling}, {Chubb}, {Min}, {Carone}, \& {Schneider}}]{samra2023}
{Samra}, D., {Helling}, C., {Chubb}, K.~L., {et~al.} 2023, \aap, 669, A142, \dodoi{10.1051/0004-6361/202244939}

\bibitem[{{Setterholm} {et~al.}(2023){Setterholm}, {Monnier}, {Le Bouquin}, {Anugu}, {Ennis}, {Jocou}, {Ibrahim}, {Kraus}, {Anderson}, {Chhabra}, {Codron}, {Farrington}, {Flores}, {Gardner}, {Gutierrez}, {Lanthermann}, {Majoinen}, {Mortimer}, {Schaefer}, {Scott}, {ten Brummelaar}, \& {Vargas}}]{setterholm2023}
{Setterholm}, B.~R., {Monnier}, J.~D., {Le Bouquin}, J.-B., {et~al.} 2023, Journal of Astronomical Telescopes, Instruments, and Systems, 9, 025006, \dodoi{10.1117/1.JATIS.9.2.025006}

\bibitem[{{Stevenson} {et~al.}(2014){Stevenson}, {D{\'e}sert}, {Line}, {Bean}, {Fortney}, {Showman}, {Kataria}, {Kreidberg}, {McCullough}, {Henry}, {Charbonneau}, {Burrows}, {Seager}, {Madhusudhan}, {Williamson}, \& {Homeier}}]{stevenson2014}
{Stevenson}, K.~B., {D{\'e}sert}, J.-M., {Line}, M.~R., {et~al.} 2014, Science, 346, 838, \dodoi{10.1126/science.1256758}

\bibitem[{{ten Brummelaar} {et~al.}(2005){ten Brummelaar}, {McAlister}, {Ridgway}, {Bagnuolo}, {Turner}, {Sturmann}, {Sturmann}, {Berger}, {Ogden}, {Cadman}, {Hartkopf}, {Hopper}, \& {Shure}}]{tenbrummelaar2005}
{ten Brummelaar}, T.~A., {McAlister}, H.~A., {Ridgway}, S.~T., {et~al.} 2005, \apj, 628, 453, \dodoi{10.1086/430729}

\bibitem[{Virtanen {et~al.}(2020)Virtanen, Gommers, Oliphant, Haberland, Reddy, Cournapeau, Burovski, Peterson, Weckesser, Bright, {van der Walt}, Brett, Wilson, Millman, Mayorov, Nelson, Jones, Kern, Larson, Carey, Polat, Feng, Moore, {VanderPlas}, Laxalde, Perktold, Cimrman, Henriksen, Quintero, Harris, Archibald, Ribeiro, Pedregosa, {van Mulbregt}, \& {SciPy 1.0 Contributors}}]{scipy}
Virtanen, P., Gommers, R., Oliphant, T.~E., {et~al.} 2020, Nature Methods, 17, 261, \dodoi{10.1038/s41592-019-0686-2}

\bibitem[{{Winterhalder} {et~al.}(2024){Winterhalder}, {Lacour}, {M{\'e}rand}, {Kammerer}, {Maire}, {Stolker}, {Pourr{\'e}}, {Babusiaux}, {Glindemann}, {Abuter}, {Amorim}, {Asensio-Torres}, {Balmer}, {Benisty}, {Berger}, {Beust}, {Blunt}, {Boccaletti}, {Bonnefoy}, {Bonnet}, {Bordoni}, {Bourdarot}, {Brandner}, {Cantalloube}, {Caselli}, {Charnay}, {Chauvin}, {Chavez}, {Choquet}, {Christiaens}, {Cl{\'e}net}, {Coud{\'e} du Foresto}, {Cridland}, {Davies}, {Dembet}, {Dexter}, {Drescher}, {Duvert}, {Eckart}, {Eisenhauer}, {F{\"o}rster Schreiber}, {Garcia}, {Garcia Lopez}, {Gardner}, {Gendron}, {Genzel}, {Gillessen}, {Girard}, {Grant}, {Haubois}, {Hei{\ss}el}, {Henning}, {Hinkley}, {Hippler}, {Houll{\'e}}, {Hubert}, {Jocou}, {Keppler}, {Kervella}, {Kreidberg}, {Kurtovic}, {Lagrange}, {Lapeyr{\`e}re}, {Le Bouquin}, {Lutz}, {Mang}, {Marleau}, {Molli{\`e}re}, {Monnier}, {Mordasini}, {Mouillet}, {Nasedkin}, {Nowak}, {Ott}, {Otten}, {Paladini}, {Paumard}, {Perraut}, {Perrin}, {Pfuhl}, {Pueyo}, {Ribeiro}, {Rickman},
  {Rustamkulov}, {Shangguan}, {Shimizu}, {Sing}, {Stadler}, {Straub}, {Straubmeier}, {Sturm}, {Tacconi}, {van Dishoeck}, {Vigan}, {Vincent}, {von Fellenberg}, {Wang}, {Widmann}, {Woillez}, \& {Yazici}}]{winterhalder2024}
{Winterhalder}, T.~O., {Lacour}, S., {M{\'e}rand}, A., {et~al.} 2024, \aap, 688, A44, \dodoi{10.1051/0004-6361/202450018}

\bibitem[{{Zhao} {et~al.}(2011){Zhao}, {Monnier}, {Che}, {Pedretti}, {Thureau}, {Schaefer}, {ten Brummelaar}, {M{\'e}rand}, {Ridgway}, {McAlister}, {Turner}, {Sturmann}, {Sturmann}, {Goldfinger}, \& {Farrington}}]{zhao2011}
{Zhao}, M., {Monnier}, J.~D., {Che}, X., {et~al.} 2011, \pasp, 123, 964, \dodoi{10.1086/661762}

\end{thebibliography}
\bibliographystyle{aasjournal}



\appendix
\begin{figure}[H]
\centering
\includegraphics[width=0.55\textwidth]{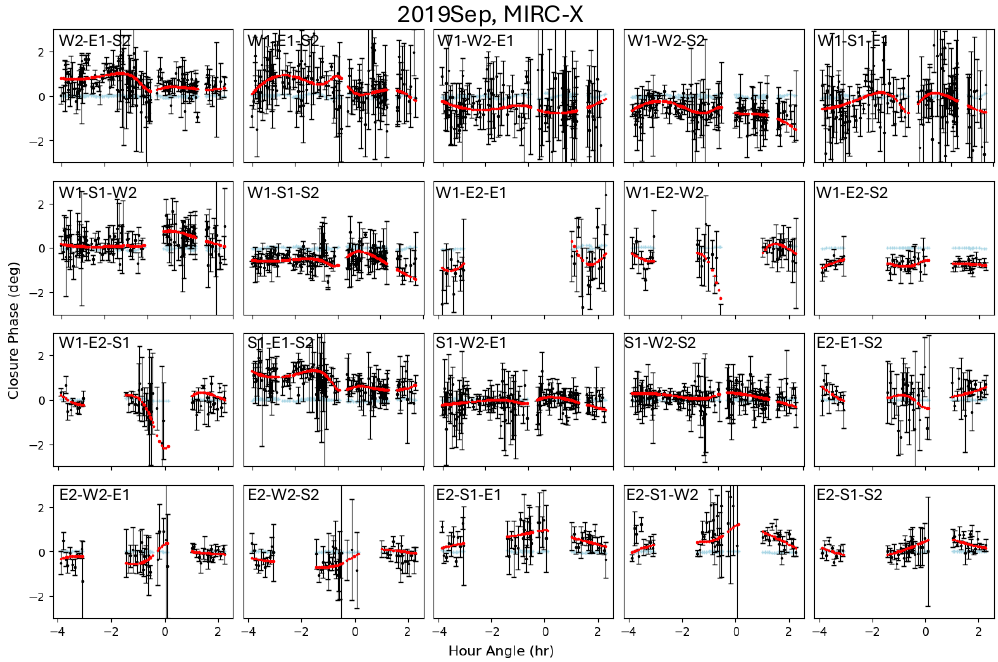}
\caption{ We plot the MIRC-X closure phase for one spectral channel across hour angle during the 2019Sep run. Each box represents one of the 20 closing triangles of the CHARA Array. The black data points represent the closure phase data, while the red points show our model of systematic trends. The light blue crosses show the expected planet signal at a planet-to-star contrast ratio of 2e-4.}
\label{trends_2019Sep}
\end{figure}

\begin{figure}[H]
\centering
\includegraphics[width=\textwidth]{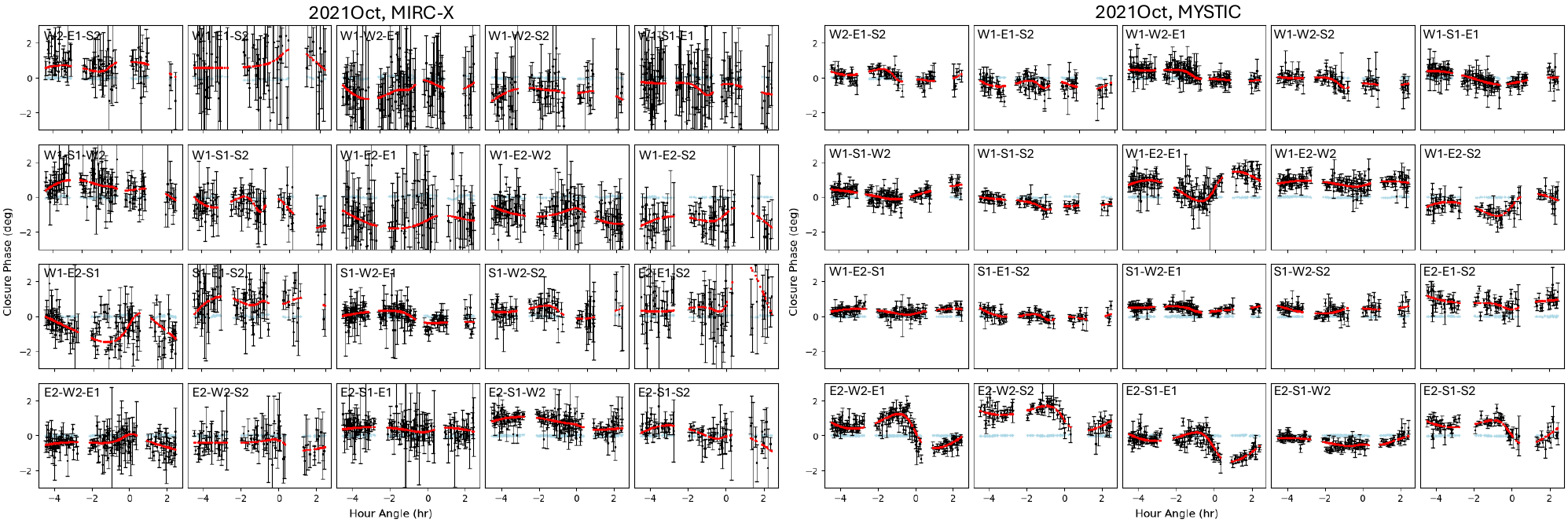}
\caption{ We plot the MIRC-X (left) and MYSTIC (right) closure phase for one spectral channel across hour angle during the 2021Oct run. Each box represents one of the 20 closing triangles of the CHARA Array. The black data points represent the closure phase data, while the red points show our model of systematic trends. The light blue crosses show the expected planet signal at a planet-to-star contrast ratio of 2e-4.}
\label{trends_2021Oct}
\end{figure}

\end{document}